# Progress on a Miniature Cold-Atom Frequency Standard

David R. Scherer, *Microsemi*; Robert Lutwak, *DARPA*; Mark Mescher, Richard Stoner, Brian Timmons,
Fran Rogomentich, Gary Tepolt, *Charles Stark Draper Laboratory*; Sven Mahnkopf, *Avo Photonics*;
Jay Noble, Sheng Chang, Dwayne Taylor, *Microsemi*

## ABSTRACT

Atomic clocks play a crucial role in timekeeping, communications, and navigation systems. Recent efforts enabled by heterogeneous MEMS integration have led to the commercial introduction of Chip-Scale Atomic Clocks (CSAC)[1] with a volume of 16 cm$^3$, power consumption of 120 mW, and instability (Allan Deviation) of σ(τ = 1 sec) < 2e-10. In order to reduce the temperature sensitivity of next-generation CSACs for timing applications, the interaction of atoms with the environment must be minimized, which can be accomplished in an architecture based on trapped, laser-cooled atoms. In this paper, we present results describing the development of a miniature cold-atom apparatus for operation as a frequency standard. Our architecture is based on laser-cooling a sample of neutral atoms in a Magneto-Optical Trap (MOT) using a conical retro-reflector in a miniature vacuum chamber. Trapping the atoms in vacuum and performing microwave interrogation in the dark reduces the temperature sensitivity compared to vapor-cell CSACs. We present details of the component development associated with the laser systems, opto-electronics, and vacuum package for miniature cold-atom technology. Finally, we conclude by characterizing the optimum alkali background pressure for such a cold-atom frequency standard.

## INTRODUCTION

Our approach is based on using a sample of laser-cooled alkali atoms created in a MOT in a miniature ultra-high vacuum (UHV) chamber. Similar related early efforts led to the observation and microwave spectroscopy of the hyperfine clock transition in laser-cooled atoms[2,3,4]. More recent efforts have focused on clock demonstration and miniaturization, including efforts based on an intracavity sample of cold atoms[5], isotropic laser cooling[6], and laser cooling in a 6-beam MOT[7]. The Miniature Cold-Atom Frequency Standard (MCAFS) effort[15,8] is based on the preparation of a laser-cooled sample in a conical retroreflector.

The clock operates by laser cooling a sample of neutral atoms in a vacuum chamber, optically pumping the sample of cold atoms into the lower hyperfine ground state, microwave population transfer of atoms from the lower to the upper hyperfine ground state, and subsequent optical state readout.

One significant advantage of the conical MOT architecture, compared to a conventional six-beam MOT, is the ability to perform laser cooling, repumping, optical state preparation, and readout with a single frequency-modulated laser beam. The laser frequency is stabilized a few linewidths on the red side of the D2 cycling transition to perform efficient Doppler cooling. Repumping is performed by modulation of the laser base current to create a sideband offset by the ground-state hyperfine frequency. The conical retro-reflector geometry eliminates the need to create six power-balanced laser beams of appropriate polarization, and the overall size of the physics package is significantly reduced.

This paper is organized as follows. The remainder of the introductory section describes the operation of the clock, the tradeoffs involved in laser cooling in a conical MOT, and requirements for laser frequency and amplitude control in such a device. The next section describes miniature cold-atom clock component development in detail for the MCAFS [85]Rb prototype. Subsequently, results describing microwave spectroscopy in a tabletop experimental setup using [133]Cs are described. Finally, our conclusions regarding miniature component development and design tradeoffs are presented.

**Theory of Operation**
A schematic showing the high-level system architecture for MCAFS is shown in Figure 1. The laser is frequency-stabilized to an atomic transition in the laser lock. The primary laser beam passes through a shutter and beam-forming optics to arrive at the vacuum package, which contains an alkali source and the conical retro-reflector. Laser cooling is performed in the MOT cone, and fluorescence from the MOT is imaged by a lens onto a photodetector for state readout.

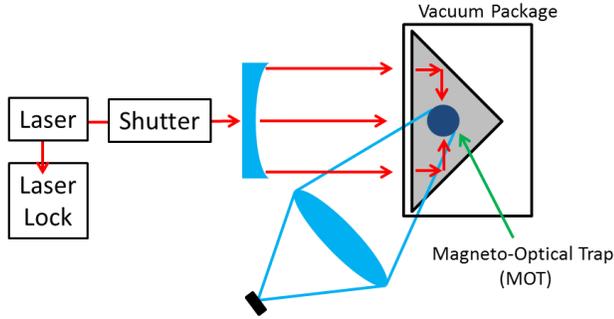

**Figure 1. System architecture for MCAFS.**

The timing sequence for pulsed microwave clock operation involves MOT loading, optical pumping, microwave interrogation, and fluorescence detection, followed by MOT re-capture and reloading. The amplitude and frequency of the laser cooling light, repump modulation, MOT magnetic field, and microwave field are varied according to the timing sequence shown in Figure 2.

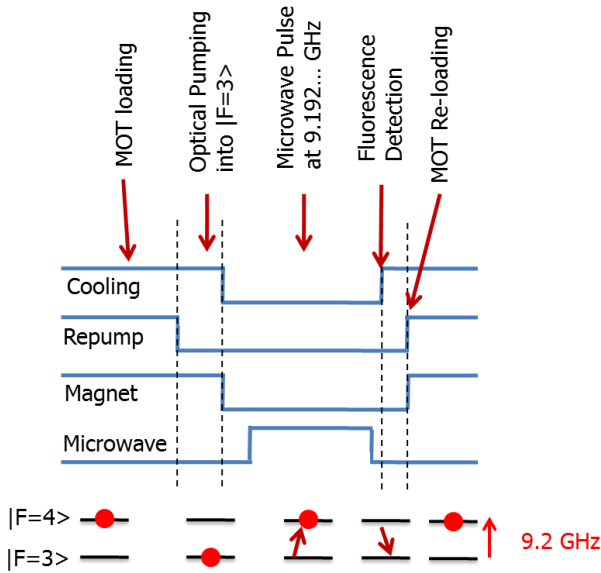

**Figure 2. Timing sequence for MCAFS operation. Also shown are energy levels for the $^{133}$Cs clock transition.**

**Laser Cooling in a Conical MOT**

Laser cooling in a pyramid or conical MOT has been performed elsewhere[9], but is not a common approach to cold-atom generation in laboratory-scale experiments. The principal reason for adopting this architecture for a miniature cold-atom clock is for miniaturization of the physics package, reduction in the total number of components, and simplification of the system architecture. The conical MOT is relatively insensitive to the alignment of the input beam, requires only one quarter-wave plate for polarization control, and requires no beamsplitters or mirrors for beam alignment.

In a traditional 6-beam MOT generated from 6 independent laser beams of diameter $D$, the steady-state number of atoms in the MOT $N_{ss} = \frac{D^2}{\sigma}\left(\frac{v_c}{u}\right)^4$, where σ is the collision cross-section with background alkali atoms, $v_c$ is the capture velocity, and $u$ is the mean thermal velocity[10]. The $D^2$ term comes from the dependence on the surface area of the laser-atom interaction region, whereas the $v_c^4$ term comes from integration of the Boltzmann velocity distribution of the background vapor. The capture velocity $v_c$ may be expressed in terms of the scattering rate $r$ and the recoil velocity $v_{rc}$ as $v_c = \sqrt{2rv_{rc}D}$, and these two expressions lead to MOT number vs. beam diameter scaling of $N_{ss} \sim D^4$ (or $D^{3.6}$ based on more detailed calculations[11]). Assuming use of a laser beam of diameter $D$ for laser cooling, the surface area of a conical chamber of outer diameter $D$ is reduced by a fraction of about 40% compared to a sphere of diameter $D$ (representative of the laser-atom interaction area in a 6-beam MOT), decreasing the eventual number of atoms in the MOT. Additionally, the number of atoms that can be slowed in a conical MOT is further reduced through the dependence of the capture velocity on the mean laser-atom interaction distance, which is smaller along many trajectories in the conical geometry. These effects combine to result in a reduced atom number in a conical MOT compared to a 6-beam MOT for a given laser beam diameter $D$. However, the overwhelming technical simplicity of the conical MOT necessitates this compromise for a miniature device.

**Requirements for Laser Frequency and Amplitude Control**

The requirements for laser frequency and amplitude control are driven by a need to minimize the light shift on the clock transition and prevent off-resonant scattering during microwave interrogation. After loading atoms in a MOT and then optical pumping into the lower hyperfine ground state, the scattering rate must be reduced. This is accomplished by reducing the beam intensity using an electro-optic switch and simultaneously increasing the detuning from the |F=3>→|F'=4> cycling transition (for $^{85}$Rb).

A complete quantum-mechanical calculation of the light shift is beyond the scope of this paper. However, a simple approximation can be derived for the case of the 3.0 GHz $^{85}$Rb ground-state |F=2> to |F=3> clock transition as follows. Considering the six dipole-allowed transitions in the $^{85}$Rb D2 manifold, one can express the light-atom interaction in terms of an arbitrary laser frequency and polarization. The light shift can be computed based on the transition dipole strengths of each transition. In the limit of large red detuning from the |F=3>→|F'=4> transition, the major contributors are the interaction of the detuned beam with the |F=3>→|F'=4>, |F=3>→|F'=3>, and |F=3>→|F'=2> transitions.

The light shift $\delta_{AC}$ can be expressed in terms of the Rabi frequency $\Omega_R$ and detuning $\Delta$ as $\delta_{AC} = \frac{\Omega_R^2}{4\Delta}$. For the power levels in MCAFS, ($I \approx I_{sat}$), $\delta_{AC} = \frac{\Omega_R^2}{4\Delta}$ = 30 kHz for $\Delta$ = -2 GHz, or, $\delta_{AC}$ = 1e-5, relative to the 3.0 GHz clock transition frequency. This implies that, for $\Delta = -2$ GHz and laser beam attenuation of 50 dB, the light shift leads to a frequency offset of 1e-10. Thus, achieving clock uncertainty below 1e-13 requires active laser intensity stabilization to one part in a thousand.

At higher scattering rates, the observed signal contrast of microwave population transfer between clock states may be degraded. This is because off-resonant scattering from |F=3> to |F'=3> or |F'=2> followed by spontaneous decay to |F=2> during the microwave interrogation time will drive spurious incoherent population transfer from |F=2> to |F=3>. The depumping rate of the |F=3> ground state population can be calculated by considering this off-resonant scattering rate into the two allowed excited state levels. This is plotted below in Figure 3 as a function of red detuning from the cycling transition and laser beam extinction. To prevent off-resonant scattering from depopulating the |F=3> state during a microwave interrogation time on the order of 20 ms, a beam extinction of 55 dB at detuning of $\Delta = -500$ MHz are required.

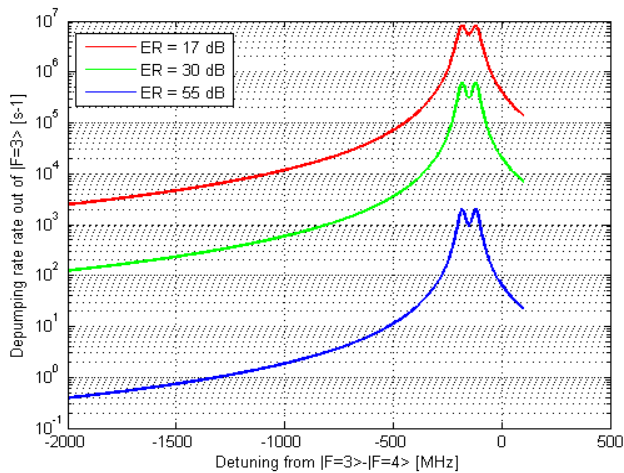

Figure 3. Depumping rate of |F=3> due to off-resonant scattering.

## MINIATURE COLD-ATOM CLOCK COMPONENT DEVELOPMENT

A miniature physics package designed for operation as a cold-atom clock based on the system architecture described above was developed. The physics package consists of an opto-electronic module housing the laser, laser lock, electro-optic shutters, and laser beam delivery optics, as well as a vacuum cell module that contains the MOT cone, vacuum chamber, magnet coils, and photodiode imaging ring. The Distributed Bragg Reflector (DBR) laser emits the majority of its light through the front facet for MOT operation and also produces a µW-level backside emission for frequency stabilization. The complete physics package is 12 cm long and occupies a total volume of 81 cm$^3$ and has been used for generation of a $^{85}$Rb MOT.

### Opto-Electronic Module

A schematic of the opto-electronic module is shown in Figure 4. The opto-electronic module generates > 5 mW of output laser power that is frequency-stabilized to ~ -2Γ from the $^{85}$Rb $^2S_{1/2}$ |F=3> → $^2P_{3/2}$ |F'=4> transition. The miniature clock physics package was designed to operate with $^{85}$Rb, rather than $^{87}$Rb or $^{133}$Cs, because the ground-state hyperfine splitting is only 3.0 GHz, compared to 6.8 GHz or 9.2 GHz, respectively. While this choice results in reduced microwave Q, $^{85}$Rb was chosen for the initial miniature physics package implementation because of the ease of FM modulation of DBR lasers at lower frequencies. Additionally, Rb has the advantage of commercial availability of miniature optical isolators at 780 nm.

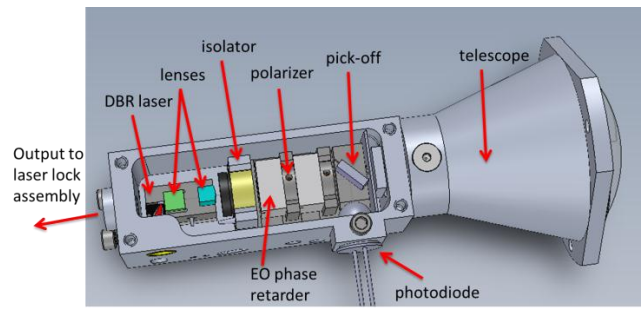

Figure 4. Schematic of Opto-Electronic Module, not including laser lock assembly.

The DBR laser is a Photodigm chip-on-submount packaged laser diode operating at a nominal wavelength of 780 nm with a threshold current of around 50 mA and slope efficiency of 0.67 W/A. The device was operated near 100 mA and packaged on a custom submount with an integrated thermo-electric cooler (TEC). The laser was frequency-modulated at a frequency of ~3.0 GHz to generate sidebands for the repump transition at |F=2> → |F'=3>. Because the laser beam must be shuttered to high extinction during microwave interrogation, a portion of the laser light must be directed to an independent laser lock assembly before the shutters. To accomplish this, a few tens of microwatts emitting from the back facet (nominally HR-coated) passes through a Thorlabs IO-D-

780-VLP Faraday optical isolator and enters the laser lock assembly.

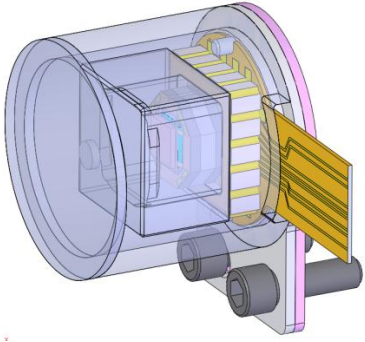

**Figure 5. Laser lock assembly.**

A drawing of the laser lock assembly is shown in Figure 5. The laser lock assembly is based on a modified physics package from a Microsemi SA.45s CSAC, using a MEMS vapor cell containing natural Rb heated by a resistive cell heater. Laser light from the back facet of the DBR passes through an aperture, a quarter-wave plate, the Rb vapor cell, and then arrives at a photodiode for laser frequency stabilization. A small permanent magnet on the back side of the Rb vapor cell provides sufficient magnetic field to shift the Doppler-broadened absorption peaks such that the side of the $^{85}$Rb |F=3> to excited state absorption feature coincides with the approximate lock point for generating a MOT.

The Doppler lock has the drawback that the lock point is set by properties of the absorption resonance which vary with the laser intensity and cell temperature. The locking position was calibrated by monitoring the saturated absorption spectrum of the DBR front-facet emission in the vacuum cell module and the laser lock spectrum of the DBR back-facet emission simultaneously, as shown in Figure 6.

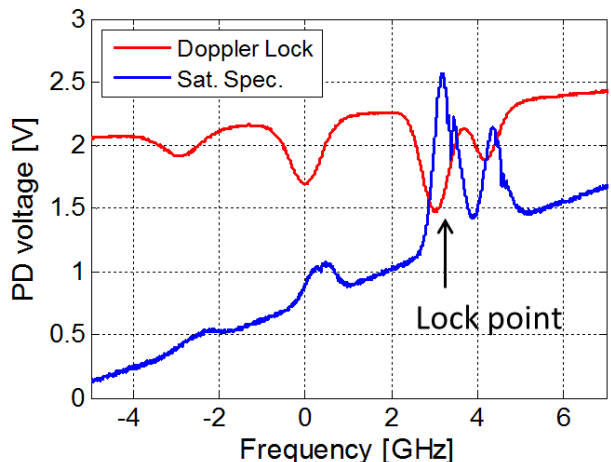

**Figure 6. Laser lock spectrum compared to saturated absorption spectroscopy (fluorescence) in the MOT cone.**

As shown in Figure 6, four Doppler-broadened peaks are visible on the laser lock spectrum, shown in red. From left to right, these four peaks correspond to the $^{87}$Rb |F=1>, $^{85}$Rb |F=2>, $^{85}$Rb |F=3>, and $^{87}$Rb |F=2> D2 lines. The two central lines correspond to $^{85}$Rb, and are separated by ~ 3.0 GHz. The fluorescence emission from laser light entering the MOT cone (when operated at a high Rb background density) is shown in blue in Figure 6. Saturated absorption dips are visible, which provide an absolute frequency calibration for the laser lock. The polarization of the light entering the laser lock assembly and the strength of the magnetic field were chosen such that the slope of the side of the absorption line (labeled 'Lock point' in Figure 6) overlaps with the $^{85}$Rb |F=3> → |F=4> transition, allowing for a dispersive error signal after subtracting a DC offset from the laser lock signal. The laser is frequency-stabilized by using this error signal in a control loop to stabilize the laser current. Dynamic corrections to the laser detuning were accomplished via corrections to this control loop. Based on the linewidth of the error signal, the maximum detuning achievable is approximately half of the Doppler width, or about ~200 MHz.

The temperature of the $^{85}$Rb laser lock cell is controlled by a resistive heater and stabilized to a set point based on a thermistor reading near the vapor cell. In our implementation, the thermal resistance between the vapor cell and the environment allows for drift on the timescale of a few seconds, which results in laser frequency drift on the order of a few MHz over a similar timescale, enough to cause visible fluctuations in the MOT fluorescence. Such fluctuations could be prevented in the future by implementation of an absolute frequency lock or corrected for by servo-locking the laser current based on the MOT signal.

The DBR front-facet emission passes through a pair of lenses to circularize and collimate the beam, and then passes through a Thorlabs IO-D-780-VLP Faraday optical isolator with a nominal 50% transmission and 40 dB isolation. After the isolator, the beam passes through a two-stage shutter assembly based on ceramic electro-optic phase retarders. The shutter configuration consists of a retarder-polarizer-retarder-polarizer stack with independent control of the two phase retarders. The phase retarders, shown in Figure 7, were Boston Applied Technologies Inc. (BATi) retarders with a 2mm x 3mm clear aperture and a specified insertion loss of 0.4 dB and extinction ratio of 30 dB at 780 nm. The active material is a quadratic ceramic electro-optic material[12] with an electric-field induced birefringence that modifies the transmission of light through a subsequent polarizer. The polarizers are Codixx colorPol VIS 041 BC6 CW02 polarizers with a specified transmission of 82% and contrast ratio of 60 dB at 780 nm. The retarders require a transverse electric field generated by a voltage on the order of 250 V in order to accomplish a 90° polarization rotation. The combination of a single stage retarder and polarizer provides an extinction ratio of ~17 dB in a response time of less than 10 μs. In our implementation, the combination of two stages of shutter assemblies does not double the extinction ratio, even with independent shutter voltage control. We speculate that this is due to light scattering or imperfections in the fidelity of the polarization rotation on a single-stage phase retarder.

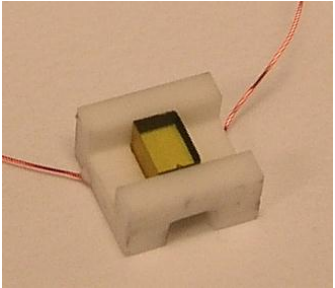

**Figure 7. EO shutter.**

Several test units were built as part of this development effort. After resolving issues associated with laser failures due to catastrophic optical damage on the DBR front facets, the only remaining functional electro-optic module operated with a single-stage shutter due to a lead failure on the second-stage shutter, resulting in a maximum of 17 dB beam extinction for the remaining work.

A beam-expanding telescope is used to enlarge the laser beam to 10 mm diameter. The electro-optic module is designed to attach to an adjustable waveplate assembly that attaches to the vacuum cell module, which creates the required circularly-polarized beam incident into the MOT cone.

**Vacuum Cell Module**

The vacuum cell module consists of a miniature vacuum chamber containing an alkali source as well as an outer housing containing an anti-Helmholtz pair of magnet coils for MOT operation and a ring of photodiodes for fluorescence collection, as shown in Figure 8. The vacuum cell is a titanium body chamber with a sapphire window and a 1/8" diameter, 6-inch long pumpout port that is connected to a Gamma Vacuum mini 0.2 L/s ion pump. The rear portion of the vacuum cell contains UHV feedthroughs connected to a SAES Rb alkali metal dispenser (AMD), which evaporates Rb into the chamber via ohmic heating. The MOT is formed in the interior of an optical-quality copper conical retro-reflector that is attached to the chamber body via a snap-ring assembly. Small apertures, or castellations, at the rim of the cone, allow for vapor-phase Rb to enter the interior of the cone, i.e. there is no direct line-of-sight between the Rb AMD and the laser cooling region.

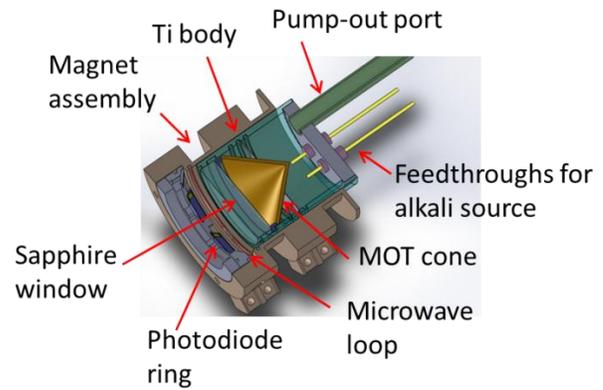

**Figure 8. Vacuum Cell Module.**

The MOT cone is formed by diamond turning an Al mandrel, which is a negative of the cone surface. Copper is then electro-chemically deposited onto the mandrel. The copper cone is subsequently removed from the mandrel, which may be re-used. A schematic of the mandrel and removed cone is shown in Figure 9. In this way, Cu cones with an apex full-angle of 90° and a conical diameter of 10 mm have been fabricated. After assembly and initial bakeout, a MOT is formed by varying the current in the Rb AMD to generate Rb background density in the interior of the MOT cone. The gauge on the ion pump, which has poor conductance to the interior of the MOT cone based on its geometry, indicates that the overall pressure in the vicinity of the ion pump is < 1e-9 Torr, independent of the AMD current.

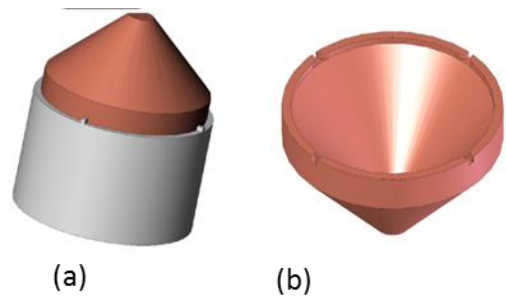

**Figure 9. Drawing of (a) mandrel and (b) copper cone after removal from mandrel.**

The vacuum cell is surrounded by a slip-on housing containing magnet coils wound in Anti-Helmholtz configuration, a ring of photodiodes for fluorescence collection, and a loop for microwave field generation. The gradient field axis is collinear with the cone axis, which is orthogonal to the direction of gravity in our laboratory. The minimum of the gradient field is positioned to be nominally in the middle of the central axis of the cone. A ring of 7 Si photodiodes wired in parallel and angled towards the center of the MOT cone form the final portion of the outer assembly. This photodiode ring is used in a non-imaging configuration for fluorescence collection of light from the interior of the MOT cone (collecting non-resonant light scatter, resonance fluorescence from the thermal alkali vapor, and resonance fluorescence from the MOT) and restricts the maximum diameter of the input laser cooling beam slightly, as shown in Figure 8. The total active area of the 7 photodiodes comprising the photodiode ring is 58 mm$^2$ and occupies a relative solid angle corresponding to about 1% of the isotropic light fluorescence from the MOT, not including multiple reflections off the cone.

**Electronics**

To the greatest extent possible, board-level electronics operate most of the components of the miniature physics package, particularly for controlling the timing sequence for MOT operation, optical pumping, and state readout. A custom electronics board based on a TI MSP430 microprocessor and Microsemi IGLOO nano FPGA control the timing of the microwave sidebands for repump modulation, the microwave coil, the MOT magnetic field, the laser frequency setpoint, and the EO shutters. A photograph of the board-level electronics is shown in Figure 10. During MOT operation, the electronics consume a total power of 2.2 W for laser cooling, optical pumping, and fluorescence detection (we note that only the laser TEC, shutter supply, and ion pump were operated on external lab power supplies), the total volume of the board is 162 cm$^3$.

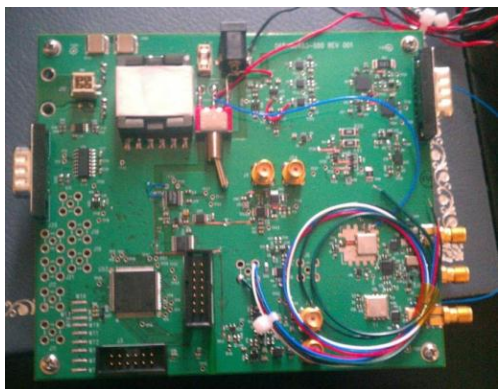

**Figure 10. Photograph of board-level electronics.**

**Component Integration and Laser Cooling in the Conical MOT**

A photograph of the complete physics package is shown in Figure 11. The entire unit is 12 cm long and occupies a total volume of 81 cm$^3$ (36 cm$^3$ for the opto-electronic module and 45 cm$^3$ for the vacuum cell module).

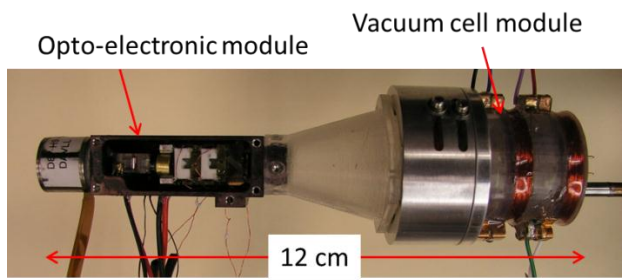

**Figure 11. Photograph of complete integrated physics package.**

To record a CCD camera image of the MOT in the system shown in Figure 11, the opto-electronic module was physically displaced from the vacuum cell module by several cm in order to allow sufficient field-of-view for an imaging camera to record the fluorescence from the interior of the MOT cone. A MOT is first observed by adjusting the alkali background pressure to a point where fluorescence is just visible, turning on a gradient magnetic field and sidebands for repump modulation, and adjusting the frequency of the laser cooling beam to a point that is a few linewidths red of the cycling transition. A camera image showing fluorescence from the MOT is shown below in Figure 12.

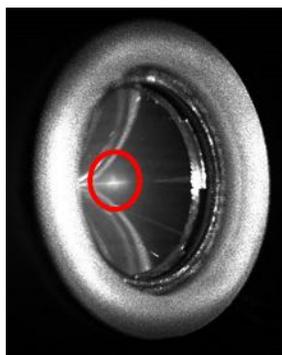

**Figure 12. Fluorescence from the MOT. The diameter of the MOT cone is 10 mm and the apex of the MOT cone is obstructed from the camera field of view by the housing on the left side of the image above. Fluorescence from the MOT is visible inside the red circle in the central portion of the interior of the MOT cone.**

Control of the Rb background density in our vacuum cell is achieved by adjustment of current through the Rb AMDs or thermal control of the chamber body via a TEC placed near the AMD feedthroughs. Current control of the Rb AMDs is used to initially coat the chamber walls with a layer of Rb and for laser frequency calibration. Final alkali vapor pressure control measurements are performed by localized heating or cooling of the vacuum cell in the vicinity of the AMD feedthroughs, which varies the Rb background density. After weeks of operation, at nominal laboratory temperature and with the Rb AMDs off, the MOT loading curve vs. time is measured by monitoring the total fluorescence signal on the photodiode ring, as shown in Figure 13.

When fit to the standard expression governing the MOT loading rate, $N(t) = N_{ss} \cdot (1 - e^{-t/\tau})$, the MOT loading time at ambient laboratory temperature (28°C as measured with a thermocouple near the Rb AMD feedthroughs on the vacuum cell) is τ = 330 ms. By heating or cooling the localized vacuum cell heater to 32°C or 19°C with the AMD current off, we can decrease or increase the MOT loading time to $\tau_{min}$ = 250 ms and $\tau_{max}$ = 620 ms, respectively. Because we are able to increase the MOT loading time (decrease the MOT loading rate) by cooling the chamber, we infer that we are operating the MOT in the regime of alkali background-limited operation, rather than the pressure-limited, or non-alkali limited, regime. We note that in a single-chamber vacuum system in which a sample of cold atoms is prepared amidst a background of thermal alkali vapor, high-SNR optical readout of the hyperfine clock states will be hindered by resonant scattering from the thermal alkali background vapor. Results from experiments intended to quantify the optimal alkali background for cold-atom clock operation in such a single-chamber system will be described subsequently.

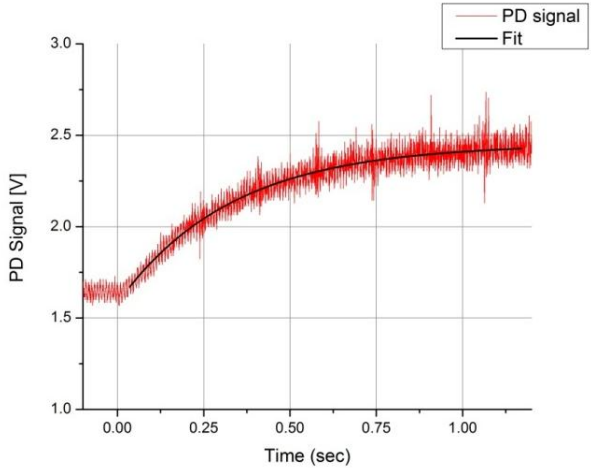

**Figure 13. MOT loading vs. time, data shown in red and fit to curve in black. Note that photodiode signal recovery electronics included the subtraction of a baseline voltage to generate this plot, the vertical axis is not meant to indicate an absolute scale.**

The miniature physics package performs cold-atom generation, optical pumping, and optical state readout after a variable microwave interrogation time. In the initial prototype, mechanical failure of one stage of the EO shutter limited our available laser beam extinction to 17 dB. Despite detuning the laser by a few hundred MHz the relatively high depumping rate at this combination of detuning and beam extinction (see Figure 3) prevented the observation of microwave population transfer between hyperfine ground states.

## MICROWAVE SPECTROSCOPY IN A TABLE-TOP EXPERIMENTAL SETUP

In order to characterize the microwave Rabi signal vs. design parameters such as alkali background pressure, a laboratory-scale experiment was constructed, utilizing $^{133}$Cs, rather than $^{85}$Rb, due to equipment availability. This table-top setup incorporated a diamond-turned 24 mm diameter MOT cone mounted inside a stainless steel UHV chamber with glass windows and a SAES Cs/NF/8/25/FT 10 + 10 alkali metal dispenser accessible through a UHV feedthrough. The chamber included an ion gauge and was pumped by a Balzers 230 L/s turbo pump backed by a roughing pump. After initial bakeout and Cs AMD activation, the chamber achieved a base pressure of 5e-10 Torr as measured on the ion gauge.

The laser system was based on a Photodigm 852 nm DBR laser that was offset-frequency-locked to a reference laser. The reference laser was locked to the Doppler-free spectrum in an optically-pumped Cs beam tube[13]. Using a heterodyne beat note, the cooling laser was capable of rapid detuning by several hundred MHz from the $^{133}$Cs D2 |F=4⟩ →|F'=5⟩ cycling transition. Repump modulation of the laser field was accomplished with a New Focus 4851 free-space phase modulator operating at ~ 9 GHz. Shuttering was accomplished via two Agiltron polarization-maintaining fiber-coupled switches connected in series followed by a free-space mechanical shutter. A beam extinction of 40 dB due to the fiber switches was accomplished in 500 ns, followed by full extinction due to the mechanical shutters after a few ms delay. A quadrupole magnetic field was generated with external magnet coils and a macroscopic imaging lens was used to image the MOT onto a photodiode. A planar double-dipole microwave antenna generates the microwave magnetic field at the location of the MOT. The double-dipole antenna contained an aperture through which the input laser beam could pass and was placed ~ 1 cm from the location of the MOT.

A camera image of the Cs MOT is shown in Figure 14. Temperature measurements based on ballistic expansion of the MOT indicate a temperature on the order of 100 μK, which is highly dependent on optimization of the x-y-z nulling coils used to position the minimum of the gradient magnetic field with respect to the overlap of the incident laser beam and conical retro-reflector structure.

In this setup, the core operation of our clock architecture, microwave spectroscopy of a sample of laser-cooled atoms prepared in a conical MOT generated by a single frequency-modulated laser beam, can be characterized in a larger laboratory-scale apparatus with superior experimental control compared to the compact physics package.

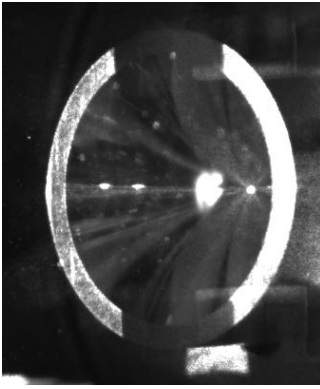

**Figure 14. Photograph of Cs MOT in laboratory-scale table-top setup. The cone is 24 mm in diameter. Scattering from the cone apex is visible as the small spot on the right in the interior of the cone, while fluorescence from the MOT is visible as the large cloud in the center of the cone.**

Microwave spectroscopy on the sample of laser-cooled atoms was demonstrating using the following timing sequence. A MOT was loaded using a cooling beam intensity of $I_{sat}$ and detuning of -1 Γ from the |F=4⟩ →|F'=5⟩ cycling transition with the repump and magnetic field gradient on. An additional cooling step of duration 5 ms with the cooling beam at a detuning of -5 Γ was performed. The sample of cold atoms was then optically pumped into the |F=3⟩ hyperfine ground state by turning the repump modulation off for 3 ms. The cooling beam is then shuttered off by using the combination of fiber switches and the mechanical shutter while the magnetic field gradient is simultaneously switched off. During the subsequent dark time, the residual magnetic bias field is controlled by additional magnet coils. The microwave field is turned on for a Rabi time of 10.3 ms, driving atoms in the magnetically-insensitive clock state from |F=3, $m_F$=0⟩ to |F=4, $m_F$=0⟩. After the microwave pulse, the cooling beam is turned back on resonance with the |F=4⟩ →|F'=5⟩ cycling transition for optical state readout. This is accomplished via integration of a small region-of-interest centered on the MOT on a CCD camera image such as shown in Figure 14. Subsequent to the microwave interrogation step, a MOT is re-loaded for the next clock cycle. We have observed an instantaneous MOT re-capture of about 50% after the dark time. A Rabi microwave spectrum is observed by varying the microwave frequency about 9.192631770 GHz based on a microwave frequency synthesizer referenced to an H-maser.

A spectrum showing a signal proportional to the number of atoms that have undergone the microwave transition vs. microwave frequency is shown in Figure 15. The horizontal axis represents frequency offset from 9.192631770 GHz, and the Rabi signal FWHM is 200 Hz. With no special effort undertaken to prepare cold atoms in any particular magnetic sublevel within the |F=3⟩ state, we expect about 1/7 of the optically-pumped atoms to be in |F=3, $m_F$=0⟩, and available to undergo microwave population transfer. The efficacy of microwave population transfer can then be compared to a control case based on optical preparation in |F=4⟩ followed by subsequent readout, accomplished by omitting the optical pumping step in the timing sequence above. This reveals that a maximum of 8% of the |F=3⟩ cold atoms undergo microwave population transfer, about one half of the expected 14%.

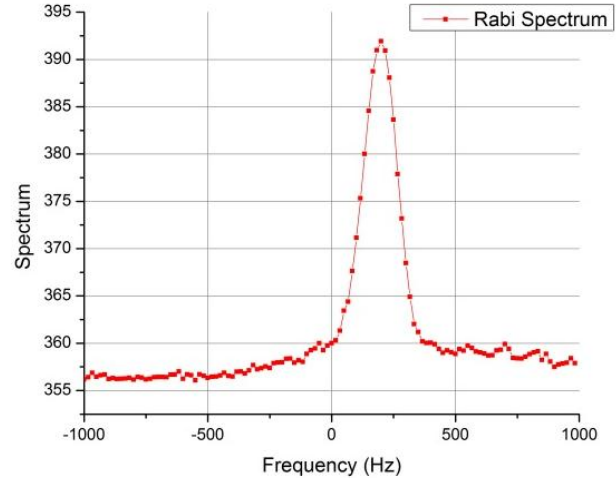

**Figure 15. Rabi spectrum of laser-cooled Cs atoms in laboratory-scale setup. The data is averaged over 28 sweeps.**

In the laboratory-scale table-top system, we are able to control the alkali background pressure via the Cs AMD while monitoring the overall chamber pressure using the ion gauge, in order to investigate the optimum alkali background pressure for clock operation. In the limit of low alkali background pressure, the MOT loading time increases, increasing the clock cycle time $T_c$ and the dead time between microwave interrogations. In the limit of high alkali background pressure, collisions decrease the number of atoms in the MOT and scattering from the thermal background vapor is increased, thereby decreasing the optical readout SNR.

These effects can be encapsulated in the standard expression for the Allan deviation of the fractional frequency fluctuations $\sigma(\tau) = \frac{1}{Q \cdot SNR}\sqrt{\frac{T_c}{\tau}}$ in the case of a pulsed atomic clock with a cycle time $T_c$ and a single-shot signal-to-noise ratio SNR. The quality factor, Q, of the measured clock transition, is limited by the microwave interrogation time and is on the order of $10^8$ for a 100 Hz linewidth. The figure-of-merit for optimizing the ADEV scales with the single-shot SNR and is inversely proportional to the square root of the clock cycle time $T_c$. In the case of thermal background-limited operation, the microwave signal SNR is proportional to the MOT-to-background fluorescence ratio collected on the fluorescence detection system, while the clock cycle time

$T_c$ ultimately scales with the MOT loading time. The clock ADEV figure-of-merit, therefore, should scale as the MOT signal-to-background ratio SBR divided by the square root of the MOT loading time $T_c$ (or, multiplied by the square root of the MOT loading rate $R$). This can be expressed as $FOM = SBR \cdot \sqrt{R}$.

To characterize this effect, we varied the current through the Cs AMD and monitored the MOT signal on the fluorescence collection photodiode, the MOT-to-background ratio SBR, the MOT loading rate $R$, and the chamber pressure as measured on the ion gauge. At each dispenser current, the chamber was allowed to equilibrate for an hour before recording a MOT loading curve. A plot of the MOT loading rate $R$ ($R = 1/\tau$, where $\tau$ = MOT loading time) vs. overall chamber pressure is show in Figure **16**. The horizontal axis is representative of the overall chamber pressure in the vicinity of the ion gauge, and should not be interpreted as the alkali background pressure in the vicinity of the MOT, which is more appropriately characterized by the MOT loading rate.

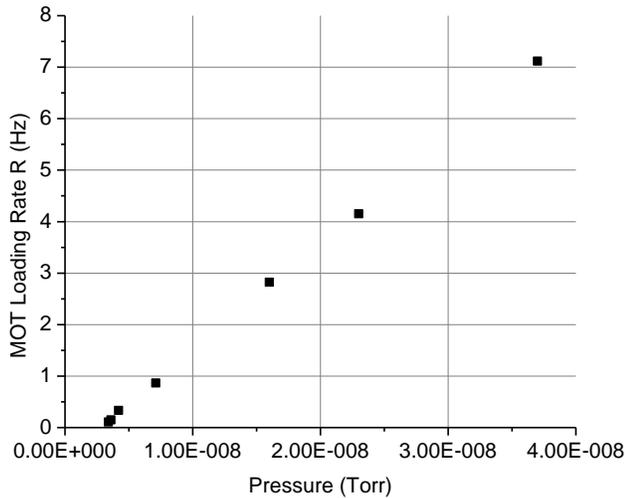

**Figure 16. MOT loading rate $R$ vs. overall chamber pressure.**

To use the MOT as a vacuum pressure gauge and calculate the overall chamber pressure based on the MOT loading time, we fit a slope to Figure 16 and observe a pressure scaling of P = (5e-9 Torr sec) · $R$, similar to within a factor of 4 of results published elsewhere[14].

The microwave clock figure-of-merit is plotted as a function of MOT loading rate $R$ in Figure 17, which reveals a maximum around $R$ = 4 Hz. For a MOT with an exponential loading rate characterized by $N(t) = N_{ss} \cdot \left(1 - e^{-t/\tau}\right)$, where $\tau$ = 250 ms, it takes 80 ms to load from 50% capacity, a typical MOT re-capture rate in our experiment, to 95% capacity. Therefore, a clock timing sequence that involves an optical pumping time of 2 ms, a microwave interrogation time of ~ 10 ms, a fluorescence detection time of 1 ms, and a MOT re-loading time of 80 ms results in a total clock cycle time on the order of $T_c$ = 100 ms, or a 10 Hz repetition rate. This is similar to the 200 ms clock cycle time reported in a previous-generation setup[15]. With a clock cycle time of $T_c$ = 100 ms, a microwave Q of $10^8$, and a single-shot SNR of 170, we expect $\sigma(\tau) = 2 \cdot 10^{-11}/\sqrt{\tau}$.

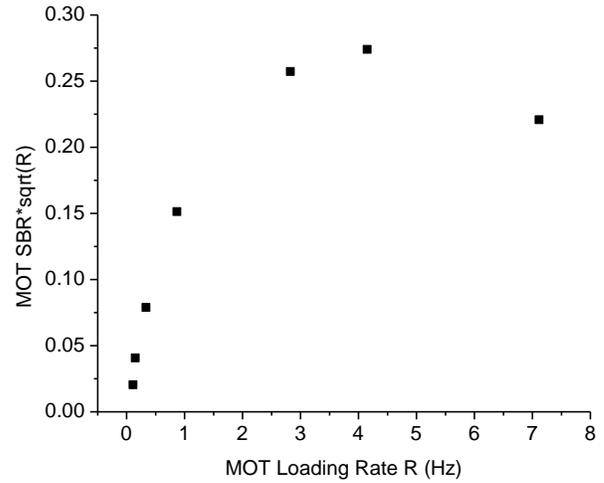

**Figure 17. Clock figure-of-merit vs. MOT loading rate $R$.**

## CONCLUSION

The MCAFS prototype resulted in a self-contained cold-atom system with a volume of 243 cm$^3$ (36 cm$^3$ opto-electronic module + 45 cm$^3$ vacuum cell module + 162 cm$^3$ control electronics) and power consumption of 2.2 W (excluding laser TEC, shutter power supply, and ion pump). To our knowledge, this is the smallest fully integrated cold-atom system (physics package volume of 81 cm$^3$) and lowest-power cold-atom system (board-level electronics consuming 2.2 W) published to date.

Finally, a characterization of the microwave spectroscopy and MOT loading vs. alkali background pressure was performed in the table-top setup, illuminating the optimal clock cycle time for future work. In a single-chamber system, there is a tradeoff between high-SNR optical readout and rapid clock cycle time, and an optimum alkali background pressure and MOT loading rate can be determined as described above.

It is our expectation that next-generation, low-power CSACs will be based on trapped atoms, and can be developed in the future using the laser-cooled neutral-atom approach as demonstrated in MCAFS. In order to develop a miniature, low-power, cold-atom CSAC, improvements in component technology are required, specifically high-efficiency laser sources, laser locking and frequency control techniques, beam shuttering, optical isolators, passive vacuum operation, alkali vapor pressure control, and low-power local oscillator technology.

## ACKNOWLEDGMENTS

We thank Richard Overstreet for valuable discussions. This research was developed with funding from the Defense Advanced Research Projects Agency (DARPA). The views, opinions, and/or findings contained in this article/presentation are those of the author(s)/presenter(s) and should not be interpreted as representing the official views or policies of the Department of Defense or the U.S. Government. Distribution Statement "A" (Approved for Public Release, Distribution Unlimited).